\documentclass[twocolumn,secnumarabic,amssymb, nobibnotes, aps, pra]{revtex4-1}

\usepackage{framed}
\usepackage{amsfonts}
\usepackage{amsmath}
\usepackage{graphicx}
\usepackage{color}

\usepackage{hyperref}

\begin{document}
\title{Localized structures formed through domain wall locking in cavity-enhanced second-harmonic generation}

\author{C. Mas Arab\'{i}$^1$, P. Parra-Rivas$^{1,2}$, T. Hansson$^3$, L. Gelens$^2$, S. Wabnitz$^{4,5,6}$, F. Leo$^1$}
\affiliation{ $^1$OPERA-Photonique CP 194/5, Universit\'e Libre de Bruxelles (ULB), Av.F.D. Roosvelt 50, B-1050 Bruxelles, Belgium \\
 $^2$Laboratory of Dynamics in Biological Systems, KU Leuven Department of Cellular and Molecular Medicine, University of Leuven, B-3000 Leuven, Belgium \\
$^3$ Department of Physics, Chemistry and Biology, Linköping University, SE-581 83 Linkoping, Sweden \\ 
$^4$Dipartimento di Ingegneria dell'Informazione, Elettronica e Telecomunicazioni, Sapienza University of Rome, Via Eudossiana 18, 00184 Rome, Italy \\
$^5$ CNR-INO, Istituto Nazionale di Ottica, Via Campi Flegrei 34, I-80078 Pozzuoli (NA), Italy \\
$^6$ Novosibirsk State University, 1 Pirogova Street, Novosibirsk 630090, Russia}

\begin{abstract}
We analyze the formation of localized structures in cavity-enhanced second-harmonic generation. We focus on the phase-matched limit, and consider that fundamental and generated waves have opposite sign of group velocity dispersion. We show that these states form due to the locking of domain walls connecting two stable homogeneous states of the system, and undergo collapsed snaking. We study the impact of temporal walk-off on the stability and dynamics of these localized states.  
\end{abstract}

\maketitle

The formation of temporal localized structures (LSs) in the context of optical resonators is attracting a lot of interest. One of the main reasons for this attention is that in the spectral domain, such states correspond to coherent optical frequency combs, whose potential applications may lead to a revolutionary breakthroughs in the fields of  metrology and spectroscopy \cite{Diddams:10}. LSs arise due to a delicate balance between nonlinearity and dispersion on the one hand, and energy gain and losses on the other hand \cite{akhmediev_dissipative_2005}.


So far, temporal LSs have been mostly generated in Kerr resonators with either anomalous or normal group velocity dispersion (GVD) \cite{leo_temporal_2010,herr_temporal_2014-2,xue_mode-locked_2015-1,garbin_experimental_2017}. However, in the past few years, their emergence in quadratic resonators has been increasingly
considered, thanks to the potential of such cavities for reaching new spectral bands, and reducing the required pump power \cite{Ulvila_OL_13,Ricciardi_PRA_15,Ricciardi_MM_2020}.


The study of LSs in quadratic resonators was first focused on spatial diffractive systems, where they arise in the plane transverse  to the propagation direction. For example, the formation of solitary waves in second-harmonic generation (SHG) \cite{etrich_pattern_1997}, or the formation of LSs through the locking of domain walls (DWs)  in optical parametric oscillators \cite{oppo_formation_1994,Staliunas_PRA_98}. 

In {\it temporal} dispersive systems, however, LSs arise along the propagation direction. In the most common case, the involved waves have a strong walk-off \cite{leo_walk-off-induced_2016}, which provides the leading  dispersion mechanism \cite{Podivilov_2020_PRA}. Nevertheless, when the waves are in opposite sides of the zero dispersion wavelength, the walk-off may vanish \cite{hansson_quadratic_2018}. Under these conditions, a large variety of LSs can emerge, which are absent otherwise  \cite{Pedro_OPO, villois_soliton_2019,parra-rivas_parametric_2020,Quadratic_solitons_singly_2020}.




Here we investigate doubly-resonant phase-matched cavity-enhanced SHG in a temporal resonator with opposite signs of GVD between the fundamental and second-harmonic fields. In this context, bright and dark LSs may form, as has been reported in \cite{hansson_quadratic_2018}. However, a complete understanding of the origin, stability and bifurcation structure of such states is still lacking. In this letter we elucidate these points, showing that the aforementioned LSs arise due to the locking of DWs in a continuous wave (cw) bistability regime and that they undergo a {\it collapsed snaking} \cite{yochelis_reciprocal_2006,Pedro_Dark}, which is preserved in the presence of small walk-off. 


In the mean-field approximation, quadratic cavities can be modeled by  the following normalized equations \cite{leo_frequency-comb_2016}: 
\begin{align}
&\partial_t A=\left[-(1+i\Delta_1)-i\eta_1\partial_x^2\right]A+iBA^*+S, \nonumber\\
&\partial_t B=\left[-(\alpha+i\Delta_2)-d\partial_x-i\eta_2\partial_x^2\right]B+iA^2,
\label{eq:1}
\end{align}
which describe the envelopes of the fundamental $A$ and the second harmonic $B$ components of the electric field. In Eq.~(\ref{eq:1}), $t$ is the slow time, $x$ is the fast time, $\eta_{1,2}$ are the normalized GVD coefficients, $d$ is the rate of walk-off between the carrier frequencies, $\alpha$ is the ratio between loss coefficients of both fields, $\Delta_{1,2}$ are the phase detunings of $A$ and $B$, respectively, and $S$ represents the external field pump.  As an example of a physical system, we choose a critically coupled lithium niobate cavity pumped along the extraordinary axis at 2707 nm. Around this wavelength, the fundamental and  second harmonic are group-velocity matched ($d=0$). The normalized parameters relate to physical quantities as follows: $\eta_1=\text{sign}(k''_1)$ and $\eta_1=k''_2/|k''_1|$, where $k''_{1,2}$ are the GVD coefficients, and at these wavelengths, $\eta_1=-1$ and $\eta_2=0.5$ respectively \cite{SMITH1976332};  $d=\Delta k'(2L/(\alpha_1|k''_1|)^{1/2}$, where $L$ is the length of the resonator that we set to $L=15$ mm, $\Delta k'$ is the walk-off  between waves, and $\alpha_1$ is the loss coefficient of the fundamental wave; $S=\sqrt{\theta P_{in}}\kappa L/\alpha^2_1$, where $\theta$  is the power transmission coefficient, $\kappa$ is the nonlinear coefficient that we set to $\kappa=$5.5 $(\sqrt{\text{W}}\text{m})^{-1}$,  and $P_{in}$  is the driving field power. We fix the phase-detunings to $\Delta_2=2\Delta_1=2\Delta$ \cite{leo_frequency-comb_2016}. We consider  equal losses for both fields ($\alpha=1$) and we set $\alpha_1=0.02$, which corresponds to a cavity with a finesse $\mathcal{F}$=160.  We apply periodic boundary conditions and set the domain size to $X$=80.



Steady state solutions of Eq. (\ref{eq:1}) satisfy the condition $\partial_t A=0$. The simplest ones are homogeneous cw states $(A_h,B_h)$, obtained by setting all derivatives to zero, leading to: 
\begin{equation}
S^2=I_A\frac{(1-2\Delta^2+I_A)^2+9\Delta^2}{1+4\Delta^2},
\end{equation}
where $I_A\equiv|A_h|^2$.
\begin{figure}
	\includegraphics[scale=1]{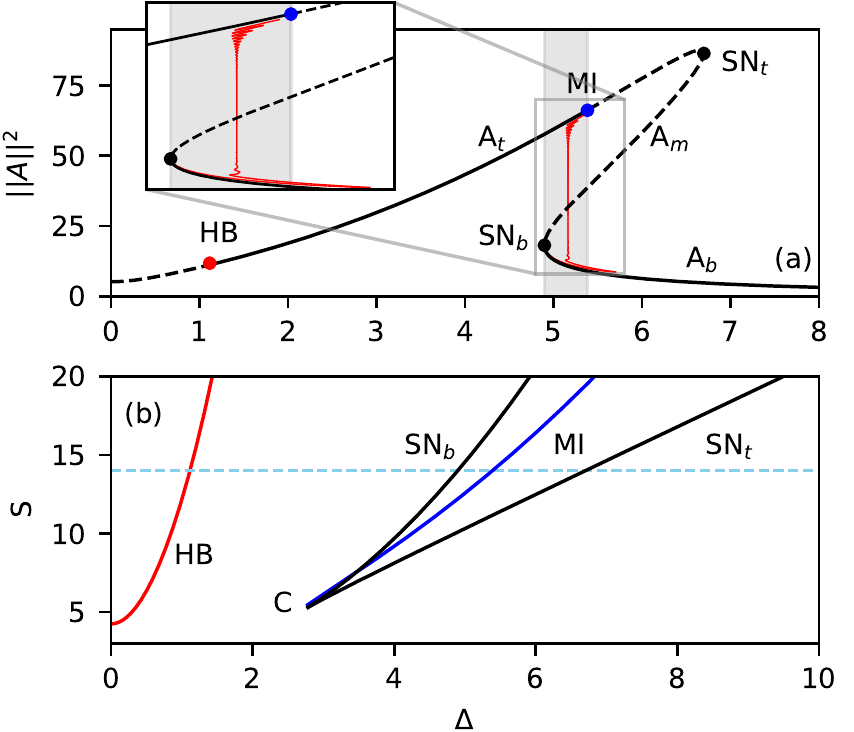}
	\caption{Panel (a) shows the cavity resonance (i.e. the cw states) for $S=14$. Solid (dashed) black line represents stable (unstable) cw state solutions. The red line represent the collapsed snaking structure associated with the LSs, and HB, SN$_{b,t}$, and MI are the Hopf, saddle-node and modulation instabilities.  Panel (b) shows the ($\Delta$, $S$)-phase diagram with the main bifurcation instability lines: the MI in blue, SN$_{b,t}$ in black and HB in red. The dashed cyan line corresponds to $S=14$. }
	\label{fig:1}
\end{figure}

\begin{figure}
		\includegraphics[scale=1]{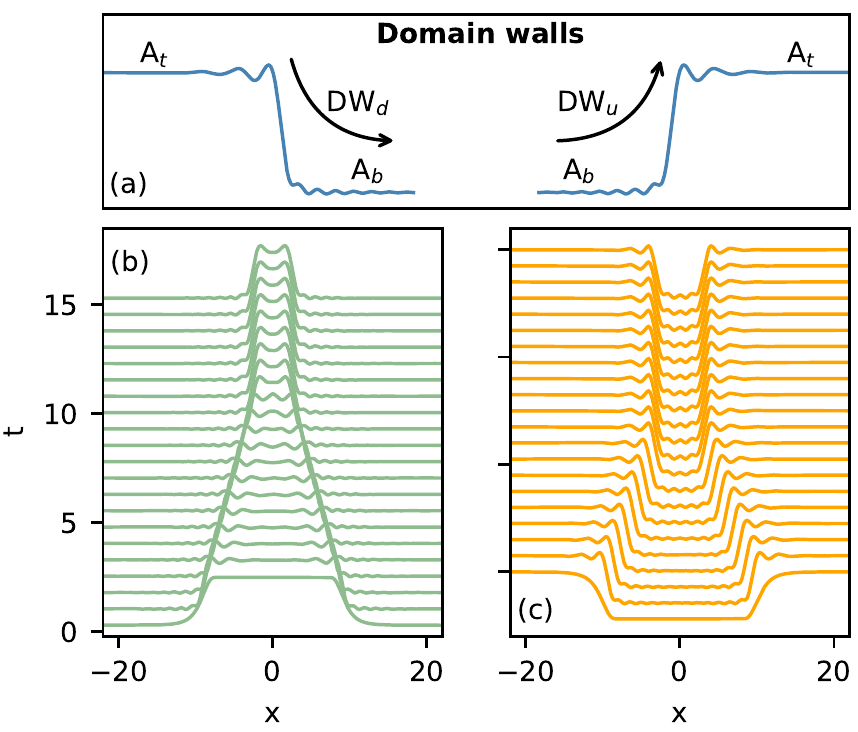}
	\caption{(a) shows the real part of DW$_d$ and DW$_u$ in the absence of walk-off ($d=0$). Panel (b) shows the formation of a bright LS through DWs locking around $A_t$. Panel (c) shows the formation of a dark state due to the locking around $A_b$. }
	\label{fig:11}
\end{figure}
Figure~\ref{fig:1}(a) shows the cw state for $S=14$ ($P_{in}=$218 mW) where we plot the $L^2$-norm of the states $||A||^2=X^{-1}\int_{-X/2}^{X/2}|A|^2dx$ as a function of $\Delta$ (i.e. the nonlinear cavity resonance).  For this particular value of $S$ the resonance is composed of three branches: $A_t$, $A_m$ and $A_b$, separated by the saddle-nodes (SN$_{b,t}$) , also referred to as folds, occurring at 
\begin{equation}
I_A^{t,b}=\frac{2(2\Delta^2-1)\pm\sqrt{4\Delta^4-31\Delta^2+1}}{3}.
\end{equation}
As a function of $\Delta$, these folds define the two bifurcation lines shown in the $(\Delta,S)-$phase diagram of Fig.~\ref{fig:1}(b), and exist whenever $4\Delta^4-31\Delta^2+1>0$. These folds disappear in a cusp bifurcation  satisfying $4\Delta^4-31\Delta^2+1=0$, and below this point the resonance becomes single valued. 
In the phase-matched limit, the steady states solutions are $\Delta\rightarrow-\Delta$ symmetric \cite{villois_soliton_2019}, and here, for simplicity, we only show the one for $\Delta>0$.

The stability of these states can be computed by studying the behavior 
of the perturbed fields $A(x,t)=A_h+a_+e^{\lambda t-ikx}+a_-e^{\lambda^* t+ik x}$, $B(x,t)=B_h+b_+e^{\lambda t-ikx}+b_-e^{\lambda^* t+ik x}$  in the linearized regime, where $\lambda$ and $k$ are the growth rate and the wavenumber of the perturbation respectively, and $a_\pm$, $b_\pm$ are small. In the absence of walk-off, this calculation leads to
\cite{trillo_pulse-train_1996,hansson_quadratic_2018}:  
\begin{equation}
\lambda=-1\pm\sqrt{-(f_1+f_2)/2\pm\sqrt{p+(f_1-f_2)^2/4}},
\end{equation}
where $f_1=\bar{\Delta}_1^2+2I_A-I_B$, $f_2=\bar{\Delta}^2_2+2I_A$, and $p=2I_A[(\bar{\Delta}_1+\bar{\Delta}_2)^2-I_B]$, with $\bar{\Delta}_1\equiv\Delta-\eta_1k^2$, $\bar{\Delta}_2\equiv2\Delta-\eta_2k^2$, and $I_B\equiv|B_h|^2$. Then the cw state $(A_h,B_h)$ is stable (resp. unstable) whenever Re$[\lambda]<0$ (resp. Re$[\lambda]>0$) for all $\lambda$. Figure~\ref{fig:1}(a) shows stable (resp. unstable) states using solid (resp. dashed) lines.

For values of $\Delta$ close to zero, $A_t$ is unstable and undergoes self-pulsing dynamics \cite{drummond_non-equilibrium_1980}. Increasing $\Delta$, $A_t$ stabilizes through a Hopf bifurcation (HB) [red dot in Fig.~\ref{fig:1}], and it remains stable until reaching modulation instability (MI) [blue dot in Fig.~\ref{fig:1}] \cite{trillo_pulse-train_1996}. At this instability, $A_t$ becomes unstable to periodic patterns, and remains so until reaching SN$_t$. In the $(\Delta,S)-$phase diagram the Hopf and MI correspond to the red and blue lines respectively [see Fig.~\ref{fig:1}(b)].
\begin{figure}
	\includegraphics[scale=1]{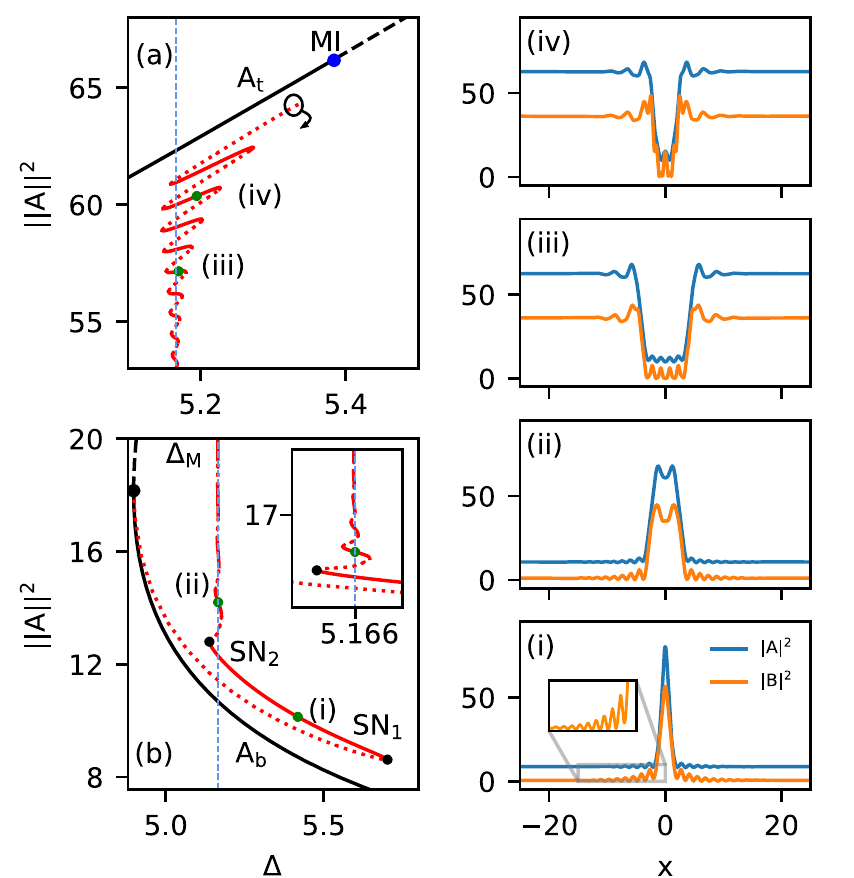}
	\caption{Panel (a) displays a close-up view of the collapsed snaking [see inset  in Fig.~\ref{fig:1}(a)] associated with dark LSs for $S=14$ and $d=0$. Panel (b) shows the collapsed snaking related to the bright states. The labels (i)-(iv) correspond to the LSs shown on the right.  }
	\label{fig:2}
\end{figure}


 Between the MI and SN$_{b}$ the system exhibits cw bistability, since $A_t$ and $A_b$ are stable and coexist [see grey-shaded region in Fig.~\ref{fig:1} (a)]. Within this region, DWs connecting $A_t$ and $A_b$  may form with two different polarities, that we label  DW$_d$ for those going from  $A_t$ to $A_b$, and DW$_u$ for connections in the opposite direction [see Fig.~\ref{fig:11}(a)]. Note that these DWs exhibit damped oscillatory tails around both $A_b$ and $A_t$, but with a different decay rate and period [see Fig.~\ref{fig:11}(a)]. 
 
 
  
 DW$_u$ and DW$_d$ drift with constant speed in opposite directions. The velocities depend on the values of the control parameters. However, for fixed pump powers, there is a value of $\Delta$ where the velocity cancels out and the DWs neither drift apart nor merge. It is referred to as the Maxwell point ($\Delta_M$) \cite{Chomaz_PRL_92}. In the neighbourhood of $\Delta_M$, the relative speed of the DWs remains small and thus, they can lock when their oscillatory tails overlap \cite{Pedro_DW,Pedro_OPO}. Indeed, the presence of oscillatory tails around $A_t$ and $A_b$ leads to locking around both homogeneous states, and therefore, to the formation of bright and dark LSs. Fig.~\ref{fig:11}(b) and (c) show an example of the time evolution and locking of two DWs leading to the formation of bright and dark LSs respectively.

  

 
In the absence of walk-off ($d=0$), these LSs undergo a bifurcation structure like the one depicted in Fig.~\ref{fig:1}(a) [see red lines], where we plot $||A||^2$ as a function of $\Delta$. The bifurcation diagrams are computed by numerically tracking the LS solutions shown in Fig.~\ref{fig:11}(b)-(c) in $\Delta$, through the free distribution software AUTO-07p \cite{Doedel07auto-07p:continuation2}. The stability of these states is calculated by computing the eigenvalues of the Jacobian matrix associated with Eq.~(\ref{eq:1}) \cite{Pedro_OPO}. 

\begin{figure}
	
	\includegraphics[scale=1]{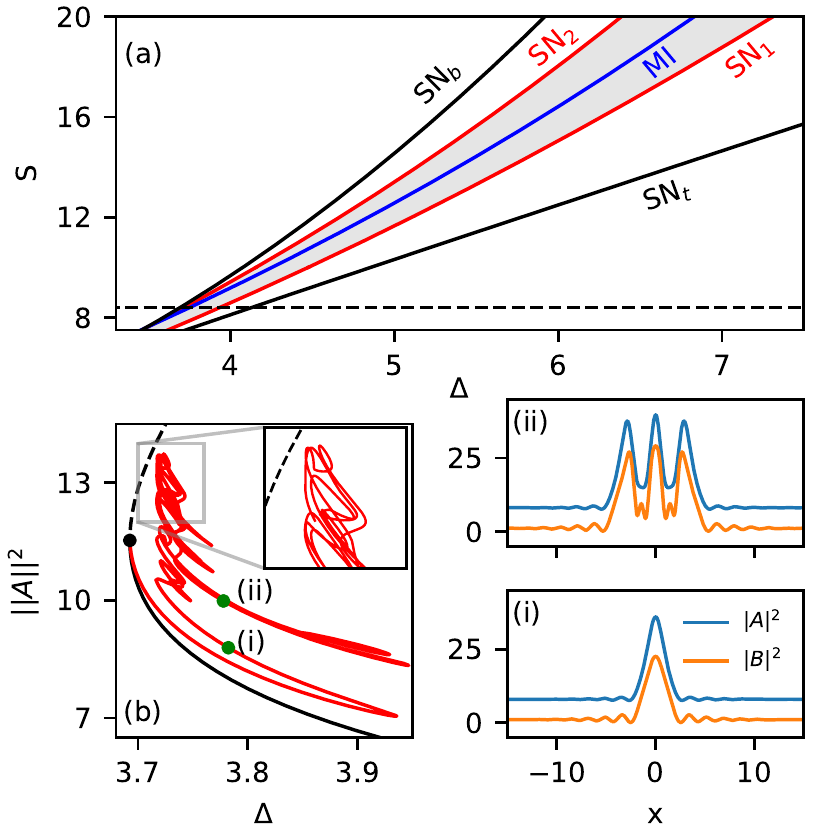}
	\caption{(a) Region of existence of LSs in the $(\Delta,S)-$phase-space, in the absence of walk-off ($d=0$), showing the main bifurcation lines of the system. (b) Bifurcation diagram for $S=8.4$ [see dashed horizontal line in panel (a)], and labels (i) and (ii) correspond to the LSs shown on the right. }
	\label{fig:3}
\end{figure}


 Figures~\ref{fig:2} (a) and (b) show a closer view of the top and bottom regions of the bifurcation diagrams plotted in Fig.~\ref{fig:1}. In both cases, the LSs undergo {\it collapsed snaking}: the LS solution branches experience a sequence of exponentially decaying oscillations around the Maxwell point $\Delta=\Delta_M$, and the different states gain and lose stability through a number of saddle-node bifurcations \cite{yochelis_reciprocal_2006,Pedro_Dark}. This bifurcation structure is a direct consequence of the DWs locking through the overlapping of their oscillatory tails \cite{Pedro_OPO}.  The difference in the forms of the tails around $A_t$ and $A_b$ [see Fig.~\ref{fig:11}(a)] leads to a different collapse snaking configurations for bright and dark states, as shown in Figs.~\ref{fig:2}(a) and (b).

Bright LSs initally arise unstably from SNb as s single peak structure. It increases its amplitude with $\Delta$, and undergoes the saddle node SN$_1$, where it becomes stable [see Fig.~\ref{fig:2}(b)]. An example of this state is shown in panel~\ref{fig:2}(i), where one can appreciate the oscillatory tails around $A_b$. Proceeding up in the diagram, the branches oscillate in $\Delta$ around $\Delta_M$, and bright LSs broaden as a period of the oscillatory tail is added at every right fold \cite{Pedro_Dark}. Figure~\ref{fig:2}(ii) shows an example of a bright state that comprises a single period of the oscillatory tails. 




 
Figure~\ref{fig:2}(a) shows the collapsed snaking diagram associated with dark LSs, such as those plotted in panels~\ref{fig:2}(iii) and (iv). In this case the branch does not terminates at the MI in $A_t$ as expected \cite{Pedro_Dark,Pedro_OPO}, but it folds back generating a complex tangle of branches, which are not shown here for simplicity [see small arrow in Fig. \ref{fig:2} (a)]. 
 
 
 
 


The region of existence of bright LSs is delimited by the first two saddle-nodes of the diagram shown in Fig.~\ref{fig:2}(b), namely SN$_{1,2}$. These two points can be tracked numerically in $\Delta$ and $S$, so that the $(\Delta,S)-$phase diagram shown in Fig.~\ref{fig:3}(a) can be computed. When increasing $S$, the region of existence of bright LSs broadens, and the collapsed snaking structure is preserved. By decreasing $S$, however, this region shrinks, and the bifurcation structure of these states becomes much more intricate, as depicted in Fig.~\ref{fig:3}(b) for $S=8.4$ ($P_{in}=78$ mW). This diagram shows a complex truss of solutions branches, whose appearance is related to the emergence of localized patterns [see Fig.~\ref{fig:3} (ii)], which form due to the heteroclinic connections occurring between the stable patterns arising from the MI and the $A_b$ state \cite{woods_heteroclinic_1999}.  The destruction of the collapsed snaking may be related to the tristablity produced by the stable periodic patterns, and the two stable cw solutions $A_b$ and $A_t$ \cite{Zelnik_2018}. 


 

  

    
\begin{figure}
	\includegraphics[scale=1]{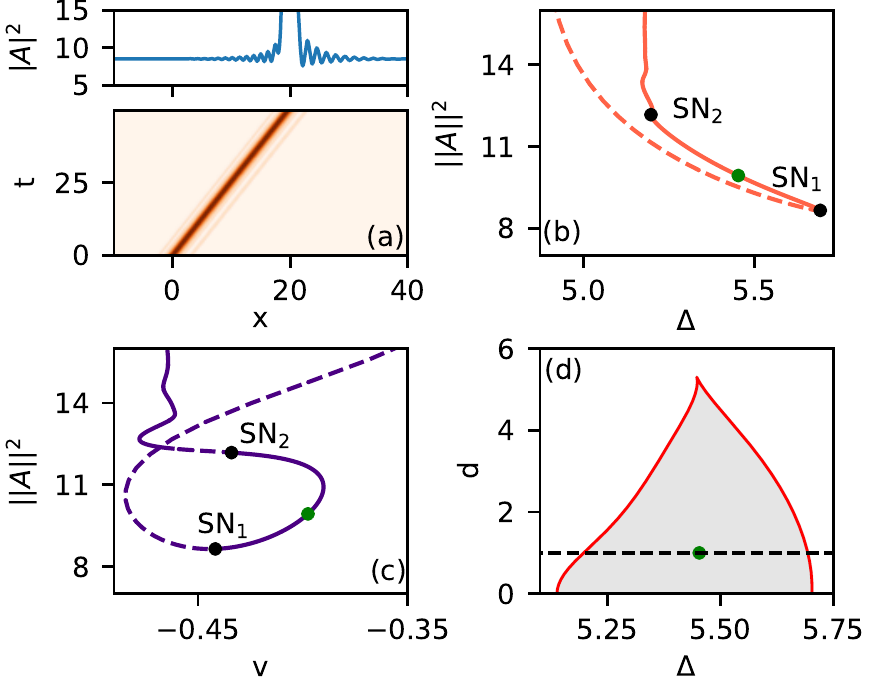}
	\caption{Panel (a) shows the temporal evolution of a bright LS in the presence of walk-off for $(\Delta,S,d)=(5.45,14,1)$. The top part of this panel shows the intensity of $A$ at $t=50$. Panel (b) shows the collapsed snaking associated with the state shown in (a), and in (c) we plot the dependence of $v$ with $||A||^2$. Panel (d) shows the $(\Delta,d)-$phase diagram for $S=14$. The black horizontal line correspond to diagrams (b) and (c). The green dots in panels (a)-(d) are related to the moving LS in (a).}
	\label{fig:4}
\end{figure}

Next, let us investigate the role of a small walk-off between the fundamental and second harmonic. The bottom panel in Fig.~\ref{fig:4}(a) shows the temporal evolution of a bright LS for the set of parameters $(\Delta,S,d)=(5.45,14,1)$. The effects of a non vanishing walk-off are twofold. On the one hand, the symmetry $x\rightarrow -x$ is broken, and as a consequence, LSs become asymmetric, as can be appreciated in the top panel of Fig.~\ref{fig:4}(a).
On the other hand, due to this asymmetry, the LS drifts with a constant speed $v$ \cite{vladimirov_effect_2018} as shown in Fig.~\ref{fig:4}(b). To numerically track these states, we recast Eq.~(\ref{eq:1}) to a moving reference frame $x'=x-vt$ where they are static, and apply parameter continuation.
Fig.~\ref{fig:4}(c) depicts the bifurcation diagram associated with the LS shown in Fig.~\ref{fig:4}(a). As we can see, it still undergo collapsed snaking. Through the numerical continuation algorithm, we are also able to  calculate the speed $v$ of these states. The output of these computations is shown in Fig.~\ref{fig:4}(c), where we plot $||A||^2$ as a function of $v$. By increasing the norm (i.e. the width of the states), the speed eventually reaches a constant value, showing that the wider LSs drift at approximately  the same speed. 

The $(\Delta,d)-$phase diagram associated with these states is plotted in Fig.~\ref{fig:4}(d) for $S=14$. The red lines represent the saddle-nodes SN$_{1,2}$ shown in Fig.~\ref{fig:4}(b), as a function of $\Delta$ and $d$. The region of existence of moving bright LSs is shadowed in light gray, and the dashed horizontal line corresponds to the diagram shown in Fig.~\ref{fig:4}(b). By increasing $d$, the region of existence of these LSs shrinks, until eventually SN$_1$ and SN$_2$ collide in a cusp bifurcation $C$ occurring at $(\Delta,d)\approx$(5.17,5.45), which in physical units corresponds to a walk-off of $\Delta k'$=2.36 ps/m. Therefore, beyond this point, single peak bright LSs are absent. Increasing $d$ further, wider LSs undergo a sequence of similar cusp bifurcations (not shown here), and this type of states finally disappears. 






In summary, we have provided a detailed bifurcation analysis of bright and dark LSs arising in dispersive cavity-enhanced second-harmonic generation with opposite signs between the fundamental and the second-harmonic waves. These states form due to the locking of DWs connecting two coexisting cw states and, for $S>10$ ($P_{in}=$ 111 mW), undergo collapsed snaking. The locking, stability, and the collapsed snaking structure result from the presence of oscillatory tails on the DWs profiles. Therefore, MI, although present, does not play an essential role in LSs formation. Furthermore, we have shown that these LSs persist for small values of the walk-off, but disappear when the walk-off is strong.


\paragraph*{Funding.}
European Research Council (ERC) under the European Union's Horizon 2020 research and innovation programe (grant agreement Nos 757800); Fonds de la Recherche Scientifique F.R.S.-FNRS; Ministry of Education and Science of the Russian Federation (14.Y26.31.0017). Swedish Research Council (Vetenskapsrådet (VR), Grant No. 2017-05309).

\bibliography{biblioteca}

\end{document}